\documentclass[aps,pra,a4paper,twocolumn,longbibliography,superscriptaddress]{revtex4-2}
\usepackage{amsmath}
\usepackage{amssymb}
\usepackage{mathtools}
\usepackage{graphicx}
\usepackage[space]{grffile}   % to include filenames with spaces
\usepackage{comment}
\usepackage{tikz}

\newcommand{\beq}{\begin{equation}}
\newcommand{\eeq}{\end{equation}}
\newcommand{\bei}{\begin{itemize}}
\newcommand{\eei}{\end{itemize}}
\newcommand{\ben}{\begin{enumerate}}
\newcommand{\een}{\end{enumerate}}

\newcommand{\be}{{\mathbf e}}

\usepackage{hyperref}
\hypersetup{
   pdfpagemode=None, 
   pdfstartpage=1,
   pdfmenubar=true,
   pdftoolbar=true,
   colorlinks = true,
   linkcolor=blue,
   citecolor=blue,
   urlcolor=blue,
   bookmarksopen=false
 }

\definecolor{darkblue}{rgb}{0.,0.24,0.51}
\definecolor{britishracinggreen}{rgb}{0.0, 0.26, 0.15}
\definecolor{darkgreen}{rgb}{0,0.60,.2}

\usepackage[normalem]{ulem}     % to strike through text using \sout{...}

\def\be{\begin{equation}}
\def\ee{\end{equation}}

\def\rf#1{(\ref{#1})}

\begin{document}
\title{Effective mass and
interaction energy of heavy Bose polarons at unitarity}
\author{Nikolay Yegovtsev}
\affiliation{Department of Physics and Center for Theory of Quantum Matter, University of Colorado, Boulder CO 80309, USA}
\author{Victor Gurarie}
\affiliation{Department of Physics and Center for Theory of Quantum Matter, University of Colorado, Boulder CO 80309, USA}

\date{\today}
	
\begin{abstract}
We study the motion of a heavy impurity immersed in a weakly interacting BEC using the Gross-Pitaevskii equation (GPe). We construct a perturbative solution to the GPe in powers of impurity velocity in the case when the boson-impurity potential is tuned to unitarity and calculate the effective mass of the polaron. In
addition, we  calculate the interaction energy of two unitary polarons which 
are sufficiently far apart. Our formalism also reproduces the results for both the mass and interaction energy obtained at weak boson-impurity coupling.

\end{abstract}
\maketitle

A Bose polaron is a quasiparticle that is formed when a quantum impurity is immersed into a gas of weakly interacting bosons below the condensation temperature. There has been a significant interest in the study of Bose polarons, especially in the regime when the impurity is allowed to interact with the bosons arbitrarily strongly \cite{Skou2021, Scazza2022, PenaArdila2019, Levinsen2021, Ardila2022, Schmidt2021, Drescher2020, Yan2020, Guenther2018, Hu2016, Jorgensen2016, Yoshida2018, Shchadilova2016, Grusdt2017}. Recently, it has been shown that when a short-ranged impurity-boson potential is tuned to a threshold of supporting a bound state, the so called unitary point, all static quasi particle properties of the Bose-polaron can be calculated analytically using the Gross-Pitaevskii equation (GPe), provided that the density of the Bose gas is sufficiently small and that the impurity-boson interactions are not too short-ranged \cite{Massignan2021,Yegovtsev2022}. In this letter we generalize these results and obtain analytic expressions for the effective mass and induced interaction energy between two unitary polarons. Our construction is based on a perturbative expansion that allows us to compute the contribution to the energy of the unitary polaron in the second order of a small parameter. First, we consider a slowly moving unitary polaron and present a derivation of the  contribution to the energy of the Bose-polaron which is quadratic in the impurity velocity.
This leads to the following induced mass
of the unitary polaron (that is, the mass of particles dragged by an impurity 
\cite{Astrakharchik2004})
\begin{equation} \label{eq:masspol}
m^*=\frac{4\sqrt{2}\pi\delta^{2/3}\xi^3n_0m}{3}.
\end{equation}
Here $m$ is the mass of the bosons
forming the condensate, $n_0$ is the density of the Bose gas  far away from the impurity, $\xi$ is the coherence length of the condensate, and $\delta=R/\xi$ where $R$ is defined below but is generally proportional to the range of the potential $r_c$. It is assumed that
$\delta \ll 1$. The calculation is done in the limit when the mass of the impurity $M \gg m$. The total effective mass of the polaron is clearly $m_{\rm polaron}=m^*+M$.

We remind the reader that the applicaiblity of the method that we use demands
$R/a_B \gg (n a_B^3)^{1/4}$, as well as $R/a_B \ll (n a_B)^{-1/2}$ 
\cite{Yegovtsev2022}. Here $a_B$
is the scattering length of the intra-boson interactons. Under these conditions, $m^* \gg m$.

Then we apply our technique to
calculate the interaction energy
of two polaron a distance $d$ apart.
We find that
\be
\label{eq:Eindst1}
E_\text{ind}(d) = -\frac{4\pi n_0 \xi^2 \delta^{2/3}}{m d}e^{-\sqrt{2}\frac{d}{\xi}}
\ee
if $d \gg \xi \delta^{1/3}$. The Yukawa form $\sim e^{-\sqrt{2}\frac{d}{\xi}}/d$ has been recently observed numerically in Ref.~\cite{drescher2023medium}.

Our calculations are valid within the same assumptions as the
calculation for the stationary polaron, that is, the potential should not be too short-ranged or too long-ranged. We show that it in order to calculate the effective mass of the polaron it is sufficient to consider only the leading term in the velocity expansion
of the solution to the GPe. Similarly, we argue that in order to compute the interaction energy between two polarons, it is sufficient to consider the solution to the GPe in the form \rf{eq:psif0}, which is the sum of two independent polaronic solutions that satisfies the boundary condition at infinity.

\label{sec:two} 
Let us now present the derivation
of these results. For the case of a moving polaron, we closely follow the method used by Astrakharchik and Pitaevskii, who solved the corresponding problem in the regime of weak boson-impurity interactions \cite{Astrakharchik2004}.
As they explained, moving polaron satisfies the equation
\begin{equation}
\label{eq:moving}
    i \frac{\partial \psi}{\partial t} = - \frac{\Delta \psi}{2m}
    +  \lambda  \left| \psi \right|^2 \psi +\left( U\left({\bf r} -
    {\boldsymbol  v} t \right)  -\mu \right) \psi.
\end{equation}
Here $U$ is the interaction potential between the polaron and the bosons,
$\lambda = 4 \pi a_B/m$, and
${\boldsymbol  v}$ is the velocity of the polaron.
The solution to this equation is of the form $\psi\left({\bf r} -
    {\bf v} t \right)$.
Substituting and changing the variables ${\bf r} -
    {\boldsymbol  v} t \rightarrow {\bf r}$,  we find
\be \label{eq:stat} - i {\boldsymbol v} \cdot \nabla \psi = - \frac{\Delta \psi}{2m}
    +  \lambda  \left| \psi \right|^2 \psi +\left( U\left({\bf r}  \right)  -\mu \right) \psi.
\end{equation}

Once the solution of the equation \rf{eq:stat} is found, the energy of the polaron
can be computed by substituting the solution into the energy
\begin{equation}
\label{eq:energy}
E = \int d^3 x \left(\frac{|\nabla \psi|^2}{2m} + \frac{\lambda}{2}|\psi|^4 + (U(\boldsymbol{r})-\mu)|\psi|^2 \right)
\end{equation}
If the impurity energy potential $U$ is weak, then Eq.~\rf{eq:stat} can be solved perturbatively in $U$. This approach was exploited by Astrakharchik and 
Pitaevski \cite{Astrakharchik2004}, who obtained the following velocity dependence of the polaron 
energy
\begin{equation}
\label{eq:APE1}
E_{v^2}  = \frac{2\sqrt{2}\pi n_0 \xi m a^2 v^2}{3},    
\end{equation}
which obviously leads to the induced mass of the weak polaron
\be \label{eq:massweak} m^*_{\rm weak} = \frac{4\sqrt{2}\pi n_0 \xi a^2 }{3} m.
\ee
Here  $a$ is the scattering length of the impurity-boson interactions. 

As the interaction strength is increased and $\left| a \right|$ grows, Eq.~\rf{eq:massweak} must break down. To calculate the induced mass of the unitary polaron whose scattering length $a$ goes to infinity, we must solve the equation
\rf{eq:stat} without assuming that $U$ is small. While we have at our disposal a technique to solve it when $U$
is tuned to unitary at ${\boldsymbol v}=0$, solving it at nonzero ${\boldsymbol v}$ appears to be difficult. Instead we propose to solve it perturbatively, although as an expansion in powers of ${\boldsymbol v}$ instead of the potential $U$. After all, we are only interested in knowing the energy \rf{eq:energy} up to
terms quadratic in velocity ${\boldsymbol v}$.

To do this we can take advantage of the following mathematical observation. Suppose we have a function $f(x,\epsilon)$, where $\epsilon$ is small. We would like to minimize $f$ with respect to $x$ and find its minimum $x_m(\epsilon)$, and we would then like to compute the value of $f$ at this minimum $f_m=f(x_m(\epsilon),\epsilon)$. We would like to calculate the expantion $f_m$  in powers of $\epsilon$  up to terms quadratic powers of $\epsilon^2$. 
We claim that to do that  it is sufficient to compute $x_m(\epsilon)$ up to terms linear in $\epsilon$. 

Indeed, suppose we compute $x_m$ up to terms quadratic in $\epsilon$. In other words,
\begin{equation} x_m = x_0 + x_1 \epsilon + x_2 \epsilon^2 + \dots.
\end{equation}
Then
it should be clear that the expansion in powers of $\epsilon$ of  $f(x_0 + x_1 \epsilon + x_2 \epsilon^2, \epsilon)$ up to terms quadratic in $\epsilon$ will not contain $x_2$, as by construction \be \left. \frac{\partial f(x,\epsilon)}{\partial x} 
\right|_{x=x_0, \, \epsilon=0 } =0.
\ee
In fact, Astrakharchik and Pitaevskii implicitly used this observation  in their paper \cite{Astrakharchik2004}.

This observation would allow us to find the energy $E$ up to terms quadratic in velocity ${\boldsymbol v}$ by
computing $\psi$ only up to terms linear in ${\boldsymbol v}$ if the equation \rf{eq:stat} that we need to solve  were obtained
as a minimization of that energy $E$. However, it is not quite so, as minimizing $E$ over $\bar \psi$ does not
produce the equation \rf{eq:stat}. Nevertheless we could write the energy \rf{eq:energy} in the following
convenient way
\begin{equation} E= E_1+ E_2,
\end{equation} where
\begin{eqnarray} E_1 &=& \int d^3 x \left(\frac{|\nabla \psi|^2}{2m} + \frac{\lambda}{2}|\psi|^4 + (U(\boldsymbol{r})-\mu)|\psi|^2 \right)+ \cr
&& i {\boldsymbol v} \int d^3 x \,  \bar \psi  \nabla \psi,
\end{eqnarray}
and
\begin{equation} E_2 =- i {\boldsymbol v} \int d^3 x \,  \bar \psi  \nabla \psi.
\ee
Minimizing $E_1$ with respect to $\bar \psi$ gives the equation \rf{eq:stat}, therefore to calculate $E_1$ up to terms quadratic in ${\boldsymbol v}$ we only need to know $\psi$ up to terms linear in ${\boldsymbol v}$ for reasons explained above. At the same time, $E_2$ is already proportional to ${\boldsymbol v}$,
therefore to calculate it up to terms quadratic in ${\boldsymbol v}$ we again need to know $\psi$ up to terms linear in ${\boldsymbol v}$. 
In other words, we can substitute $\psi$ calculated up to terms linear in 
${\boldsymbol v}$ directly into the energy \rf{eq:energy} and
find it up to terms quadratic in ${\boldsymbol v}$.
This observation significantly simplifies the required algebra. 

Let us now calculate the expansion of $\psi$ in powers of 
${\boldsymbol v}$. It is convenient to introduce dimensionless variable $\phi = \psi/\sqrt{n_0}$. It is normalized so that far away from the impurity $\phi=1$.
We then write
\be \phi = \phi_0 + \phi_1 + \dots,
\ee where $\phi_0$ is independent of ${\boldsymbol v}$, $\phi_1$ is linear in ${\boldsymbol v}$ and so on. Substituting into the equation \rf{eq:stat} and expanding in powers of ${\boldsymbol v}$ we find, first of all,
\be \label{eq:stati}- \frac{\Delta \phi_0}{2m}
    +  \mu  \left( \phi_0^2-1 \right) \phi_0 +U\left({\bf r}\right) \phi_0 = 0.
\end{equation}
Here we used that $\lambda n_0 = \mu$. We also used that $\phi_0$ is real. This
becomes clear if we note that the equation \rf{eq:stati} coincides with the 
GPe for the stationary polaron. Therefore, $\phi_0$ coincides with the 
solution of the stationary unitary polaron problem found in Refs.~\cite{Massignan2021,Yegovtsev2022}, which was real. 

At the same time, we find that
\begin{equation} \label{eq:first} - i {\boldsymbol v}  \nabla \phi_0 = - \frac{\Delta \phi_1}{2m} 
+\mu \left( \phi_0^2 - 1 \right) \phi_1 + U({\bf r}) \phi_1.
\end{equation} Here we used that $\phi_1$ as should be clear from this
expression, is purely imaginary. 
If we find $\phi_1$ by solving this equation, we will know $\psi$ up to terms
linear in ${\boldsymbol v}$. We can now substitute this into the expression for the energy \rf{eq:energy} and expand in powers of ${\boldsymbol v}$ up to
terms quadratic in it.

The zero order term is the energy of the stationary polaron. The first order term
vanishes, as it turns out. This is not surprising as those terms if they were
not zero would depend on the direction of ${\boldsymbol v}$, while we work with a rotationally invariant polaron. Finally, the terms quadratic in ${\boldsymbol v}$ can be brought to the form
\be \label{eq:ense} E_{v^2} = -i n_0 {\boldsymbol v} \int d^3 x \, \bar \phi_1 \nabla \phi_0.
\ee
This remarkably simple expression tells us that to compute the energy 
we need to use $\phi_0$ calculated in Refs.~\cite{Massignan2021,Yegovtsev2022}, determine $\phi_1$ by solving Eq.~\rf{eq:first} and substitute into the expression
\rf{eq:ense}.

Let us now review the structure of the solutions $\phi_0$ of the GPe both at weak coupling and at unitarity as described in Refs.~\cite{Massignan2021,Yegovtsev2022}.
For simplicity, we consider the potentials that vanish identically beyond some range $r_c$. The solution to Eq.~\rf{eq:stati} for the case of the weak potential $|a|^3 \ll \xi^2 r_c$, where $\xi^2 =1/(2m\mu)$ is the square of the coherence length $\xi$, reads: 

\be \label{eq:fullsolw}  \phi_0(r)  \approx \left\{  \begin{matrix}  \left( 1 - \frac a {r_c} \right)  r_c \Psi_0 \left(r \right), & r < r_c , \cr 1 - \frac a r  \exp \left( {- {\sqrt{2} r}/{\xi}} \right),  & r>r_c .   \end{matrix}
\right. \ee
Here  $\Psi_0(r)$ is the solution to the zero energy Schrodinger equation
\begin{equation}
    -\frac{1}{2m} \Delta \Psi_0 + U({\boldsymbol r})
    \Psi_0 = 0
\end{equation}
in the potential $U$ with the normalization that satisfies $\Psi_0\left( r_c \right)=1/r_c$, and $a$ is the corresponding scattering length.

When potential $U$ is tuned to unitarity, the result becomes $(\delta = {R}/{\xi} \sim r_c/\xi \ll 1)$:
\be \label{eq:fullsols}  \phi_0(r)  \approx \left\{  \begin{matrix}  \xi \delta^{1/3}   \Psi_0 \left(r \right), & r < r_c , \cr 1+\frac{\xi \delta^{1/3} }{r}   \exp \left( {-{\sqrt{2} r}/{\xi}} \right),  & r>r_c .   \end{matrix}
\right. \ee

Here $R$ is defined as $R^{-1} = \int_0^\infty dr \, r^2 \Psi_0^4$. One expects \cite{Yegovtsev2022} $R\sim r_c$. Note that in the region where $r>r_c$ both solutions have the same structure, but differ only by a coefficient in front of the Yukawa tail. 

Let us focus on the unitary case and solve
the equation \rf{eq:first} for $\phi_1(r)$. We seek solution in the form $\phi_1\left( {\boldsymbol r} 
\right)= i v P_1(\cos(\theta)) u_1(r)/r$, where $P_1 \left( \cos(\theta) \right)$ is the first Legendre polynomial, $\theta$ is the angle between ${\boldsymbol v}$ and ${\bf r}$, while $u_1(r)$ is the radial part of the solution. Plugging this form into Eq.~\rf{eq:first} we get:
\begin{equation}
\label{eq:crr} - r m \phi_0' = -\frac{u_1''}{2}  +
\frac{\phi_0^2-1}{2 \xi^2} u_1 +
 \frac {u_1} {r^2} + m U(r) u.
\end{equation}    
If the potential $U$ is unitary, 
we make the observation that  $(\phi_0^2-1)/(2 \xi^2) \ll 1/r^2$ for all $r$ and can be neglected. To see that, we note that for $r<r_c$,
$\phi_0 \sim 1/\delta^{2/3}\gg 1$ and we find
$$ (\phi_0^2-1)/(2 \xi^2) \sim 1/(\xi^2 \delta^{4/3}) \ll 1/r_c^2 < 1/r^2.$$ 
Next, for $r_c<r<\xi^{2/3} r_c^{1/3}$, 
$\phi_0 \sim \xi \delta^{1/3}/r \gg 1$ and we
find
$$(\phi_0^2-1)/(2 \xi^2) \sim \delta^{2/3}/r^2 \ll 1/r^2.$$
For $\xi^{2/3} r_c^{1/3} < r <\xi$, $\phi_0-1
\sim \xi \delta^{1/3}/r \ll 1$
and we find
$$(\phi_0^2-1)/(2 \xi^2) \sim \delta^{1/3}/(\xi r) \ll 1/r^2.$$ Finally, for $r>\xi$, $\phi_0^2-1$ decays exponentially towards zero, so it is obviously much smaller than $1/r^2$.

Note that a potential weaker than unitary 
leads to even smaller $\phi_0^2-1$, therefore, these arguments also apply at weak potential. 

These observations allow us to significantly
simplify the equation \rf{eq:crr}, with the result
\begin{equation}
\label{eq:Un}
 2mr\phi_0' = u_1'' - \frac{2}{r^2}u_1 -2mUu_1.
\end{equation}
Now for $r>r_c$, $U=0$. While $U \not = 0$ at $r<r_c$, it turns out that the
region $0<r<r_c$ does not significantly contribute to the energy and therefore
we can simplify this even further and solve
\be \label{eq:wws} 2mr\phi_0' = u_1'' - \frac{2}{r^2}u_1.
\ee
We will come back to this point below. 

The equation \rf{eq:wws} is simple enough where
it can be solved explicitly, with the solution
\begin{equation}
\label{eq:Unu1}
u_1 = \frac{2m}3 \left(  r^2 \left(\phi_0(r)
- 1 \right)  - \frac{1}{r}\int_0^rdr' r'^3 \phi_0'(r') \right).  
\end{equation}
Plugging this expression for $u_1$ together with the definition of $\phi_0(r)$,   Eq.~\rf{eq:fullsols}, into Eq.~\rf{eq:ense}, we get:
\begin{equation}
\label{eq:Ev2un}
E_{v^2} = \frac{2\sqrt{2}\pi\delta^{2/3}\xi^3n_0mv^2}{3}.    
\end{equation} The induced mass of the polaron at unitarity \rf{eq:masspol} follows immediately. 

We still need to estimate the contribution
of the region $0<r<r_c$ to the energy. In that region, $\phi_0\sim 1/\delta^{2/3}$, $u_1 \sim r_c^2/\delta^{2/3}$, so
we can find that this region contributes
$n_0 m v^2 \xi r_c^2$ to the energy. This is much smaller than \rf{eq:Ev2un} as
$\xi r_c^2 \ll \xi^3 \delta^{2/3}$.

As was discussed previously, 
both weak and strong polaron solutions
Eq.\rf{eq:fullsolw} and Eq.~\rf{eq:fullsols} have the same $r$
dependence in the region $r>r_c$.
Therefore, an obvious substitution $\delta^{1/3}\xi \to \left|a\right|$ reduces
our result for the energy of the 
unitary polaron to the energy obtained
by Astrakharchik and Pitaevskii
Eq.~\rf{eq:APE1} for the weak polaron. 

Just as the result for the energy
of the stationary polaron found
earlier in Refs.~\cite{Massignan2021,Yegovtsev2022},
the result found here Eq.~\rf{eq:Ev2un} is only valid when $\delta \ll 1$. In principle corrections to it proportional
to higher powers of $\delta$ can also be calculated if needed. 

%%%%%%%%%%%%%%%%%%%%%%%%%%%%%%%%%%%%%%%%%%%%%%%%%%%%%%%%%%%%%%%
%\section{Induced interactions}
We can also calculate the 
interaction energy of two stationary unitary polarons  separated by a distance $d$. For the weak polaron, the real space expression has been obtained before by \cite{Camacho_Guardian2018a,Camacho_Guardian2018b, Jager_2022}, and it reads:
\be
\label{eq:Eind1}
E_\text{int}(d) = -\frac{4 \pi n_0 a^2 }{ m d } e^{-\sqrt{2}\frac{d}{\xi}}
\ee

To calculate this for the unitary heavy polarons, we need to solve the GPe
with two potentials
\be \label{eq:stat2}  - \frac{\Delta \psi}{2m}
    +  \lambda  \left| \psi \right|^2 \psi +\left( U\left({\boldsymbol r} \right) +
    U\left({\boldsymbol  r}- {\boldsymbol d}\right)  -\mu \right) \psi =0
\end{equation}
and calculate the energy of the solution
\begin{eqnarray}
\label{eq:energy2}
E &= &   \int d^3 x \Biggl( \frac{|\nabla \psi|^2}{2m} + \frac{\lambda}{2}|\psi|^4 +  \cr && (U(\boldsymbol{r})+
U\left({\boldsymbol r} - {\boldsymbol d} \right) -\mu)|\psi|^2 \Biggr).
\end{eqnarray}
We will not attempt to do this for a generic separation between the polarons $d$. Let us just compute this in case when
$d$ is so large that the solution can be written as 
\be \label{eq:gpdi} \frac{\psi}{\sqrt{n_0}} \approx  \phi_0({\boldsymbol r}) +
\phi_0\left({\boldsymbol r} - {\boldsymbol d} \right) - 1
+ f\left({\boldsymbol r} \right),
\ee
where $\left| f \right| \ll 1$.
Here $\phi_0({\boldsymbol r})$
is the solution for a single
stationary unitary polaron 
\rf{eq:fullsols}. Crucially, 
$\left| f \right|$ is indeed small if 
 $\left|\phi(d)-1 \right| \ll 1$.
 This is because at very large separation between the polarons clearly $f=0$ should solve the corresponding GPe \rf{eq:stat2}. If $d$ is large but finite, $f$ will be nonzero but small. This is guaranteed by
 $\phi(d)$ approaching $1$ at $d \gg \xi \delta^{1/3}$. All of this
 can be verified by a direct substitution of \rf{eq:gpdi} into \rf{eq:stat2}
 with the help of \rf{eq:fullsols}. 

The same theorem that earlier allowed us
to compute energy for a moving polaron up to terms quadratic in velocity while calculating $\psi$ up to terms linear in it allows now
to calculate the energy of two
polarons by 
 substituting the solution for two 
 polarons at $f=0$ 
 \be \label{eq:psif0} \psi = \sqrt{n_0} \left( \phi_0({\boldsymbol r}) +
\phi_0\left({\boldsymbol r} - {\boldsymbol d} \right) - 1
\right)
\ee
into the expression \rf{eq:energy2}. Subtracting the part of the energy independent of $d$, which
is the energy of the condensate and the individual energies of the polarons, gives
the interaction energy of the polarons.

This program gives  the leading contribution to the interaction energy at large $d$. To compute corrections to that if needed, we would have to solve for $f$ by substituting Eq.~\rf{eq:gpdi} 
in the GPe \rf{eq:stat2}. We will not attempt to do it here. 

Carrying out this program produces the
interaction energy \rf{eq:Eindst1} of two polarons distance $d$ apart. Note that
replacing $\xi \delta^{1/3} \rightarrow \left|a\right|$
gives the interaction energy of weak polarons \rf{eq:Eind1}, as we should have expected on
general grounds discussed earlier. This result  is in agreement with the recent numerical study by \cite{drescher2023medium} who observed the Yukawa-type behaviour consistent with Eq.~\rf{eq:Eindst1} holding up to distances of the order of $\xi$. At smaller distances two polarons start to have a significant overlap and one cannot longer use Eq.~\rf{eq:psif0} to compute the energy. As a starting point one needs to solve the GPe in the spherically nonsymmetric potential analytically. This goes beyond formalism described in \cite{Massignan2021, Yegovtsev2022}.

As a side note, while for the unitary case we had to use the assumption that two polarons must be well separated in order to be able to use Eq.~\rf{eq:psif0}, for weak polarons this form is correct for arbitrary separation $d$. Indeed, for the weak polaron, one can linearize the GPe and the solution to the two polaron problem  will be just a linear combination of the solutions to a single polaron problem. The contribution to the interaction energy reads:
\be
\label{eq:Eindweak}
E_\text{int}(d) = 2n_0\int d^3 x \, U(r) (\phi_0({\boldsymbol r} - {\boldsymbol d})-1).
\ee
For the short-ranged potentials $r_c \ll \xi$ and for distances $d\gg r_c$, $\phi_0(\boldsymbol{r}-\boldsymbol{d}) = 1-\frac{a}{|\boldsymbol{r}-\boldsymbol{d}|}e^{-\sqrt{2}\frac{|\boldsymbol{r}-\boldsymbol{d}|}{\xi}} \approx 1-\frac{a}{d}e^{-\sqrt{2}\frac{d}{\xi}}$ in the vicinity of $r=r_c$. Recalling the definition of the scattering length that is valid in the weak coupling regime $\frac{2\pi a}{m} = \int d^3 x \, U(r)$, one retrieves Eq.~\rf{eq:Eind1}. The result in Eq.~\rf{eq:Eindweak} is valid for arbitrary distances between two impurities and arbitrary potentials conforming to the second Born approximation. For example, when $d=0$, one would retrieve physics of a single polaron sitting in the potential with the strength that is twice the strength of the original potential, provided one redefines the scattering length in an appropriate manner. In general case, the form of the induced potential will be complicated and can resemble the form of the original potential like in the case of the exponential potential $U \sim e^{-\frac{r}{r_c}}$, or the polarization potential analyzed in \cite{Ding2022} which was studied by the means of many body perturbation theory.
%Just as in calculating the velocity dependence of the energy, the contribution of the core regions of the polarons where $0<r<r_c$ and $\left| {\boldsymbol d} - {\boldsymbol r} \right| < r_c$ can be neglected.

Finally, we note that to the leading order both the effective mass and the two polaron interaction energy at unitarity are related to the weak coupling limit by the identification $\xi\delta^{1/3} \to |a|$. The same identification also reproduces results for some other quasiparticle properties, such as Tan's contact and the quasiparticle residue that has been studied in the context of the bosonic orthogonality catastrophe in Ref. \cite{Guenther2020}, but not the energy (to work for the energy, the substitution has to be modified to $\xi\delta^{1/3} \to 2|a|/3$)  \cite{Massignan2021, Yegovtsev2022}. We leave the question of how robust this property is and what other quantities obey this identification to a future study.

In summary, we derived a compact formula for the effective mass of the slowly moving impurity in the BEC which is valid both for weak and strong potentials, generalizing the result by Astrakharchik and Pitaevskii. We used it to find an analytic expression for the effective mass of the Bose polaron at unitarity. We also derived the expression for the interaction energy between two unitary polarons separated by a large distance $d$. We observed that the qualitative difference between weak and strong polarons in those scenarios can be captured by a trivial substitution of the amplitudes of the solution of the unitary polaron into the the corresponding expression at weak coupling similar to other quasiparticle properties. When applied to weak polarons, our method allows to compute the interaction energy between two polarons at arbitrary distance between two impurities and it serves as a generalization of previous results at weak coupling. Because our method relies on a perturbative construction, we were unable to compute the drag force and the induced interactions at small impurity separations which requires full nonperturbative dependence on the impurity velocity and knowledge of the solution to the GPe in the non central potential. We leave those problems for a future study.
\label{sec:three}

%\vspace{5mm}
%\begin{acknowledgments}
We acknowledge inspiring and insightful discussions with 
P. Massignan and G. Astrakharchik. 
Some of this work was carried out while the authors visited
the Universitat Polit\`ecnica de Catalunya, Spain.
%J. Levinsen, and M. Parish.
%PM was supported by grant PID2020-113565GB-C21 funded by MCIN/AEI/10.13039/501100011033, by EU FEDER Quantumcat, and by the {\it ICREA Academia} program.
 This work was also supported by the Simons Collaboration on Ultra-Quantum Matter, which is a grant from the Simons Foundation (651440, VG, NY).
%\end{acknowledgments}

\bibliography{UnitaryPolaron}

\begin{thebibliography}{23}%
\makeatletter
\providecommand \@ifxundefined [1]{%
 \@ifx{#1\undefined}
}%
\providecommand \@ifnum [1]{%
 \ifnum #1\expandafter \@firstoftwo
 \else \expandafter \@secondoftwo
 \fi
}%
\providecommand \@ifx [1]{%
 \ifx #1\expandafter \@firstoftwo
 \else \expandafter \@secondoftwo
 \fi
}%
\providecommand \natexlab [1]{#1}%
\providecommand \enquote  [1]{``#1''}%
\providecommand \bibnamefont  [1]{#1}%
\providecommand \bibfnamefont [1]{#1}%
\providecommand \citenamefont [1]{#1}%
\providecommand \href@noop [0]{\@secondoftwo}%
\providecommand \href [0]{\begingroup \@sanitize@url \@href}%
\providecommand \@href[1]{\@@startlink{#1}\@@href}%
\providecommand \@@href[1]{\endgroup#1\@@endlink}%
\providecommand \@sanitize@url [0]{\catcode `\\12\catcode `\$12\catcode
  `\&12\catcode `\#12\catcode `\^12\catcode `\_12\catcode `\%12\relax}%
\providecommand \@@startlink[1]{}%
\providecommand \@@endlink[0]{}%
\providecommand \url  [0]{\begingroup\@sanitize@url \@url }%
\providecommand \@url [1]{\endgroup\@href {#1}{\urlprefix }}%
\providecommand \urlprefix  [0]{URL }%
\providecommand \Eprint [0]{\href }%
\providecommand \doibase [0]{http://dx.doi.org/}%
\providecommand \selectlanguage [0]{\@gobble}%
\providecommand \bibinfo  [0]{\@secondoftwo}%
\providecommand \bibfield  [0]{\@secondoftwo}%
\providecommand \translation [1]{[#1]}%
\providecommand \BibitemOpen [0]{}%
\providecommand \bibitemStop [0]{}%
\providecommand \bibitemNoStop [0]{.\EOS\space}%
\providecommand \EOS [0]{\spacefactor3000\relax}%
\providecommand \BibitemShut  [1]{\csname bibitem#1\endcsname}%
\let\auto@bib@innerbib\@empty
%</preamble>
\bibitem [{\citenamefont {Skou}\ \emph {et~al.}(2021)\citenamefont {Skou},
  \citenamefont {Skov}, \citenamefont {J{\o}rgensen}, \citenamefont {Nielsen},
  \citenamefont {Camacho-Guardian}, \citenamefont {Pohl}, \citenamefont
  {Bruun},\ and\ \citenamefont {Arlt}}]{Skou2021}%
  \BibitemOpen
  \bibfield  {author} {\bibinfo {author} {\bibfnamefont {M.~G.}\ \bibnamefont
  {Skou}}, \bibinfo {author} {\bibfnamefont {T.~G.}\ \bibnamefont {Skov}},
  \bibinfo {author} {\bibfnamefont {N.~B.}\ \bibnamefont {J{\o}rgensen}},
  \bibinfo {author} {\bibfnamefont {K.~K.}\ \bibnamefont {Nielsen}}, \bibinfo
  {author} {\bibfnamefont {A.}~\bibnamefont {Camacho-Guardian}}, \bibinfo
  {author} {\bibfnamefont {T.}~\bibnamefont {Pohl}}, \bibinfo {author}
  {\bibfnamefont {G.~M.}\ \bibnamefont {Bruun}}, \ and\ \bibinfo {author}
  {\bibfnamefont {J.~J.}\ \bibnamefont {Arlt}},\ }\bibfield  {title} {\bibinfo
  {title} {\emph {{Non-equilibrium quantum dynamics and formation of the Bose
  polaron}}},\ }\href {\doibase 10.1038/s41567-021-01184-5} {\bibfield
  {journal} {\bibinfo  {journal} {Nat. Phys.}\ }\textbf {\bibinfo {volume}
  {17}},\ \bibinfo {pages} {731} (\bibinfo {year} {2021})}\BibitemShut
  {NoStop}%
\bibitem [{\citenamefont {Scazza}\ \emph {et~al.}(2022)\citenamefont {Scazza},
  \citenamefont {Zaccanti}, \citenamefont {Massignan}, \citenamefont {Parish},\
  and\ \citenamefont {Levinsen}}]{Scazza2022}%
  \BibitemOpen
  \bibfield  {author} {\bibinfo {author} {\bibfnamefont {F.}~\bibnamefont
  {Scazza}}, \bibinfo {author} {\bibfnamefont {M.}~\bibnamefont {Zaccanti}},
  \bibinfo {author} {\bibfnamefont {P.}~\bibnamefont {Massignan}}, \bibinfo
  {author} {\bibfnamefont {M.~M.}\ \bibnamefont {Parish}}, \ and\ \bibinfo
  {author} {\bibfnamefont {J.}~\bibnamefont {Levinsen}},\ }\bibfield  {title}
  {\bibinfo {title} {\emph {Repulsive Fermi and Bose Polarons in Quantum
  Gases}},\ }\href {\doibase 10.3390/atoms10020055} {\bibfield  {journal}
  {\bibinfo  {journal} {Atoms}\ }\textbf {\bibinfo {volume} {10}},\ \bibinfo
  {pages} {55} (\bibinfo {year} {2022})}\BibitemShut {NoStop}%
\bibitem [{\citenamefont {Pe\~na Ardila}\ \emph {et~al.}(2019)\citenamefont
  {Pe\~na Ardila}, \citenamefont {J\o{}rgensen}, \citenamefont {Pohl},
  \citenamefont {Giorgini}, \citenamefont {Bruun},\ and\ \citenamefont
  {Arlt}}]{PenaArdila2019}%
  \BibitemOpen
  \bibfield  {author} {\bibinfo {author} {\bibfnamefont {L.~A.}\ \bibnamefont
  {Pe\~na Ardila}}, \bibinfo {author} {\bibfnamefont {N.~B.}\ \bibnamefont
  {J\o{}rgensen}}, \bibinfo {author} {\bibfnamefont {T.}~\bibnamefont {Pohl}},
  \bibinfo {author} {\bibfnamefont {S.}~\bibnamefont {Giorgini}}, \bibinfo
  {author} {\bibfnamefont {G.~M.}\ \bibnamefont {Bruun}}, \ and\ \bibinfo
  {author} {\bibfnamefont {J.~J.}\ \bibnamefont {Arlt}},\ }\bibfield  {title}
  {\bibinfo {title} {\emph {Analyzing a Bose polaron across resonant
  interactions}},\ }\href {\doibase 10.1103/PhysRevA.99.063607} {\bibfield
  {journal} {\bibinfo  {journal} {Phys. Rev. A}\ }\textbf {\bibinfo {volume}
  {99}},\ \bibinfo {pages} {063607} (\bibinfo {year} {2019})}\BibitemShut
  {NoStop}%
\bibitem [{\citenamefont {Levinsen}\ \emph {et~al.}(2021)\citenamefont
  {Levinsen}, \citenamefont {Ardila}, \citenamefont {Yoshida},\ and\
  \citenamefont {Parish}}]{Levinsen2021}%
  \BibitemOpen
  \bibfield  {author} {\bibinfo {author} {\bibfnamefont {J.}~\bibnamefont
  {Levinsen}}, \bibinfo {author} {\bibfnamefont {L.~A.~P.}\ \bibnamefont
  {Ardila}}, \bibinfo {author} {\bibfnamefont {S.~M.}\ \bibnamefont {Yoshida}},
  \ and\ \bibinfo {author} {\bibfnamefont {M.~M.}\ \bibnamefont {Parish}},\
  }\bibfield  {title} {\bibinfo {title} {\emph {Quantum Behavior of a Heavy
  Impurity Strongly Coupled to a Bose Gas}},\ }\href {\doibase
  10.1103/PhysRevLett.127.033401} {\bibfield  {journal} {\bibinfo  {journal}
  {Phys. Rev. Lett.}\ }\textbf {\bibinfo {volume} {127}},\ \bibinfo {pages}
  {033401} (\bibinfo {year} {2021})}\BibitemShut {NoStop}%
\bibitem [{\citenamefont {Ardila}(2022)}]{Ardila2022}%
  \BibitemOpen
  \bibfield  {author} {\bibinfo {author} {\bibfnamefont {L.~A.~P.}\
  \bibnamefont {Ardila}},\ }\bibfield  {title} {\bibinfo {title} {\emph
  {Ultra-Dilute Gas of Polarons in a Bose-Einstein Condensate}},\ }\href
  {\doibase 10.3390/atoms10010029} {\bibfield  {journal} {\bibinfo  {journal}
  {Atoms}\ }\textbf {\bibinfo {volume} {10}},\ \bibinfo {pages} {29} (\bibinfo
  {year} {2022})}\BibitemShut {NoStop}%
\bibitem [{\citenamefont {Schmidt}\ and\ \citenamefont
  {Enss}(2021)}]{Schmidt2021}%
  \BibitemOpen
  \bibfield  {author} {\bibinfo {author} {\bibfnamefont {R.}~\bibnamefont
  {Schmidt}}\ and\ \bibinfo {author} {\bibfnamefont {T.}~\bibnamefont {Enss}},\
  }\bibfield  {title} {\bibinfo {title} {\emph {{Self-stabilized Bose
  polarons}}},\ }\href {\doibase 10.48550/arXiv.2102.13616} {\bibfield
  {journal} {\bibinfo  {journal} {arXiv:2102.13616}\ } (\bibinfo {year}
  {2021}),\ 10.48550/arXiv.2102.13616}\BibitemShut {NoStop}%
\bibitem [{\citenamefont {Drescher}\ \emph {et~al.}(2020)\citenamefont
  {Drescher}, \citenamefont {Salmhofer},\ and\ \citenamefont
  {Enss}}]{Drescher2020}%
  \BibitemOpen
  \bibfield  {author} {\bibinfo {author} {\bibfnamefont {M.}~\bibnamefont
  {Drescher}}, \bibinfo {author} {\bibfnamefont {M.}~\bibnamefont {Salmhofer}},
  \ and\ \bibinfo {author} {\bibfnamefont {T.}~\bibnamefont {Enss}},\
  }\bibfield  {title} {\bibinfo {title} {\emph {Theory of a resonantly
  interacting impurity in a Bose-Einstein condensate}},\ }\href {\doibase
  10.1103/PhysRevResearch.2.032011} {\bibfield  {journal} {\bibinfo  {journal}
  {Phys. Rev. Research}\ }\textbf {\bibinfo {volume} {2}},\ \bibinfo {pages}
  {032011} (\bibinfo {year} {2020})}\BibitemShut {NoStop}%
\bibitem [{\citenamefont {Yan}\ \emph {et~al.}(2020)\citenamefont {Yan},
  \citenamefont {Ni}, \citenamefont {Robens},\ and\ \citenamefont
  {Zwierlein}}]{Yan2020}%
  \BibitemOpen
  \bibfield  {author} {\bibinfo {author} {\bibfnamefont {Z.~Z.}\ \bibnamefont
  {Yan}}, \bibinfo {author} {\bibfnamefont {Y.}~\bibnamefont {Ni}}, \bibinfo
  {author} {\bibfnamefont {C.}~\bibnamefont {Robens}}, \ and\ \bibinfo {author}
  {\bibfnamefont {M.~W.}\ \bibnamefont {Zwierlein}},\ }\bibfield  {title}
  {\bibinfo {title} {\emph {Bose polarons near quantum criticality}},\ }\href
  {\doibase 10.1126/science.aax5850} {\bibfield  {journal} {\bibinfo  {journal}
  {Science}\ }\textbf {\bibinfo {volume} {368}},\ \bibinfo {pages} {190}
  (\bibinfo {year} {2020})}\BibitemShut {NoStop}%
\bibitem [{\citenamefont {Guenther}\ \emph {et~al.}(2018)\citenamefont
  {Guenther}, \citenamefont {Massignan}, \citenamefont {Lewenstein},\ and\
  \citenamefont {Bruun}}]{Guenther2018}%
  \BibitemOpen
  \bibfield  {author} {\bibinfo {author} {\bibfnamefont {N.-E.}\ \bibnamefont
  {Guenther}}, \bibinfo {author} {\bibfnamefont {P.}~\bibnamefont {Massignan}},
  \bibinfo {author} {\bibfnamefont {M.}~\bibnamefont {Lewenstein}}, \ and\
  \bibinfo {author} {\bibfnamefont {G.~M.}\ \bibnamefont {Bruun}},\ }\bibfield
  {title} {\bibinfo {title} {\emph {Bose Polarons at Finite Temperature and
  Strong Coupling}},\ }\href {\doibase 10.1103/PhysRevLett.120.050405}
  {\bibfield  {journal} {\bibinfo  {journal} {Phys. Rev. Lett.}\ }\textbf
  {\bibinfo {volume} {120}},\ \bibinfo {pages} {050405} (\bibinfo {year}
  {2018})}\BibitemShut {NoStop}%
\bibitem [{\citenamefont {Hu}\ \emph {et~al.}(2016)\citenamefont {Hu},
  \citenamefont {Van~de Graaff}, \citenamefont {Kedar}, \citenamefont {Corson},
  \citenamefont {Cornell},\ and\ \citenamefont {Jin}}]{Hu2016}%
  \BibitemOpen
  \bibfield  {author} {\bibinfo {author} {\bibfnamefont {M.-G.}\ \bibnamefont
  {Hu}}, \bibinfo {author} {\bibfnamefont {M.~J.}\ \bibnamefont {Van~de
  Graaff}}, \bibinfo {author} {\bibfnamefont {D.}~\bibnamefont {Kedar}},
  \bibinfo {author} {\bibfnamefont {J.~P.}\ \bibnamefont {Corson}}, \bibinfo
  {author} {\bibfnamefont {E.~A.}\ \bibnamefont {Cornell}}, \ and\ \bibinfo
  {author} {\bibfnamefont {D.~S.}\ \bibnamefont {Jin}},\ }\bibfield  {title}
  {\bibinfo {title} {\emph {Bose Polarons in the Strongly Interacting
  Regime}},\ }\href {\doibase 10.1103/PhysRevLett.117.055301} {\bibfield
  {journal} {\bibinfo  {journal} {Phys. Rev. Lett.}\ }\textbf {\bibinfo
  {volume} {117}},\ \bibinfo {pages} {055301} (\bibinfo {year}
  {2016})}\BibitemShut {NoStop}%
\bibitem [{\citenamefont {J\o{}rgensen}\ \emph {et~al.}(2016)\citenamefont
  {J\o{}rgensen}, \citenamefont {Wacker}, \citenamefont {Skalmstang},
  \citenamefont {Parish}, \citenamefont {Levinsen}, \citenamefont
  {Christensen}, \citenamefont {Bruun},\ and\ \citenamefont
  {Arlt}}]{Jorgensen2016}%
  \BibitemOpen
  \bibfield  {author} {\bibinfo {author} {\bibfnamefont {N.~B.}\ \bibnamefont
  {J\o{}rgensen}}, \bibinfo {author} {\bibfnamefont {L.}~\bibnamefont
  {Wacker}}, \bibinfo {author} {\bibfnamefont {K.~T.}\ \bibnamefont
  {Skalmstang}}, \bibinfo {author} {\bibfnamefont {M.~M.}\ \bibnamefont
  {Parish}}, \bibinfo {author} {\bibfnamefont {J.}~\bibnamefont {Levinsen}},
  \bibinfo {author} {\bibfnamefont {R.~S.}\ \bibnamefont {Christensen}},
  \bibinfo {author} {\bibfnamefont {G.~M.}\ \bibnamefont {Bruun}}, \ and\
  \bibinfo {author} {\bibfnamefont {J.~J.}\ \bibnamefont {Arlt}},\ }\bibfield
  {title} {\bibinfo {title} {\emph {Observation of Attractive and Repulsive
  Polarons in a Bose-Einstein Condensate}},\ }\href {\doibase
  10.1103/PhysRevLett.117.055302} {\bibfield  {journal} {\bibinfo  {journal}
  {Phys. Rev. Lett.}\ }\textbf {\bibinfo {volume} {117}},\ \bibinfo {pages}
  {055302} (\bibinfo {year} {2016})}\BibitemShut {NoStop}%
\bibitem [{\citenamefont {Yoshida}\ \emph {et~al.}(2018)\citenamefont
  {Yoshida}, \citenamefont {Endo}, \citenamefont {Levinsen},\ and\
  \citenamefont {Parish}}]{Yoshida2018}%
  \BibitemOpen
  \bibfield  {author} {\bibinfo {author} {\bibfnamefont {S.~M.}\ \bibnamefont
  {Yoshida}}, \bibinfo {author} {\bibfnamefont {S.}~\bibnamefont {Endo}},
  \bibinfo {author} {\bibfnamefont {J.}~\bibnamefont {Levinsen}}, \ and\
  \bibinfo {author} {\bibfnamefont {M.~M.}\ \bibnamefont {Parish}},\ }\bibfield
   {title} {\bibinfo {title} {\emph {Universality of an Impurity in a
  Bose-Einstein Condensate}},\ }\href {\doibase 10.1103/PhysRevX.8.011024}
  {\bibfield  {journal} {\bibinfo  {journal} {Phys. Rev. X}\ }\textbf {\bibinfo
  {volume} {8}},\ \bibinfo {pages} {011024} (\bibinfo {year}
  {2018})}\BibitemShut {NoStop}%
\bibitem [{\citenamefont {Shchadilova}\ \emph {et~al.}(2016)\citenamefont
  {Shchadilova}, \citenamefont {Schmidt}, \citenamefont {Grusdt},\ and\
  \citenamefont {Demler}}]{Shchadilova2016}%
  \BibitemOpen
  \bibfield  {author} {\bibinfo {author} {\bibfnamefont {Y.~E.}\ \bibnamefont
  {Shchadilova}}, \bibinfo {author} {\bibfnamefont {R.}~\bibnamefont
  {Schmidt}}, \bibinfo {author} {\bibfnamefont {F.}~\bibnamefont {Grusdt}}, \
  and\ \bibinfo {author} {\bibfnamefont {E.}~\bibnamefont {Demler}},\
  }\bibfield  {title} {\bibinfo {title} {\emph {Quantum Dynamics of Ultracold
  Bose Polarons}},\ }\href {\doibase 10.1103/PhysRevLett.117.113002} {\bibfield
   {journal} {\bibinfo  {journal} {Phys. Rev. Lett.}\ }\textbf {\bibinfo
  {volume} {117}},\ \bibinfo {pages} {113002} (\bibinfo {year}
  {2016})}\BibitemShut {NoStop}%
\bibitem [{\citenamefont {Grusdt}\ \emph {et~al.}(2017)\citenamefont {Grusdt},
  \citenamefont {Schmidt}, \citenamefont {Shchadilova},\ and\ \citenamefont
  {Demler}}]{Grusdt2017}%
  \BibitemOpen
  \bibfield  {author} {\bibinfo {author} {\bibfnamefont {F.}~\bibnamefont
  {Grusdt}}, \bibinfo {author} {\bibfnamefont {R.}~\bibnamefont {Schmidt}},
  \bibinfo {author} {\bibfnamefont {Y.~E.}\ \bibnamefont {Shchadilova}}, \ and\
  \bibinfo {author} {\bibfnamefont {E.}~\bibnamefont {Demler}},\ }\bibfield
  {title} {\bibinfo {title} {\emph {Strong-coupling Bose polarons in a
  Bose-Einstein condensate}},\ }\href {\doibase 10.1103/PhysRevA.96.013607}
  {\bibfield  {journal} {\bibinfo  {journal} {Phys. Rev. A}\ }\textbf {\bibinfo
  {volume} {96}},\ \bibinfo {pages} {013607} (\bibinfo {year}
  {2017})}\BibitemShut {NoStop}%
\bibitem [{\citenamefont {Massignan}\ \emph {et~al.}(2021)\citenamefont
  {Massignan}, \citenamefont {Yegovtsev},\ and\ \citenamefont
  {Gurarie}}]{Massignan2021}%
  \BibitemOpen
  \bibfield  {author} {\bibinfo {author} {\bibfnamefont {P.}~\bibnamefont
  {Massignan}}, \bibinfo {author} {\bibfnamefont {N.}~\bibnamefont
  {Yegovtsev}}, \ and\ \bibinfo {author} {\bibfnamefont {V.}~\bibnamefont
  {Gurarie}},\ }\bibfield  {title} {\bibinfo {title} {\emph {Universal Aspects
  of a Strongly Interacting Impurity in a Dilute Bose Condensate}},\ }\href
  {\doibase 10.1103/PhysRevLett.126.123403} {\bibfield  {journal} {\bibinfo
  {journal} {Phys. Rev. Lett.}\ }\textbf {\bibinfo {volume} {126}},\ \bibinfo
  {pages} {123403} (\bibinfo {year} {2021})}\BibitemShut {NoStop}%
\bibitem [{\citenamefont {Yegovtsev}\ \emph {et~al.}(2022)\citenamefont
  {Yegovtsev}, \citenamefont {Massignan},\ and\ \citenamefont
  {Gurarie}}]{Yegovtsev2022}%
  \BibitemOpen
  \bibfield  {author} {\bibinfo {author} {\bibfnamefont {N.}~\bibnamefont
  {Yegovtsev}}, \bibinfo {author} {\bibfnamefont {P.}~\bibnamefont
  {Massignan}}, \ and\ \bibinfo {author} {\bibfnamefont {V.}~\bibnamefont
  {Gurarie}},\ }\bibfield  {title} {\bibinfo {title} {\emph {Strongly
  interacting impurities in a dilute Bose condensate}},\ }\href {\doibase
  10.1103/PhysRevA.106.033305} {\bibfield  {journal} {\bibinfo  {journal}
  {Phys. Rev. A}\ }\textbf {\bibinfo {volume} {106}},\ \bibinfo {pages}
  {033305} (\bibinfo {year} {2022})}\BibitemShut {NoStop}%
\bibitem [{\citenamefont {Astrakharchik}\ and\ \citenamefont
  {Pitaevskii}(2004)}]{Astrakharchik2004}%
  \BibitemOpen
  \bibfield  {author} {\bibinfo {author} {\bibfnamefont {G.~E.}\ \bibnamefont
  {Astrakharchik}}\ and\ \bibinfo {author} {\bibfnamefont {L.~P.}\ \bibnamefont
  {Pitaevskii}},\ }\bibfield  {title} {\bibinfo {title} {\emph {Motion of a
  heavy impurity through a Bose-Einstein condensate}},\ }\href {\doibase
  10.1103/PhysRevA.70.013608} {\bibfield  {journal} {\bibinfo  {journal} {Phys.
  Rev. A}\ }\textbf {\bibinfo {volume} {70}},\ \bibinfo {pages} {013608}
  (\bibinfo {year} {2004})}\BibitemShut {NoStop}%
\bibitem [{\citenamefont {Drescher}\ \emph {et~al.}(2023)\citenamefont
  {Drescher}, \citenamefont {Salmhofer},\ and\ \citenamefont
  {Enss}}]{drescher2023medium}%
  \BibitemOpen
  \bibfield  {author} {\bibinfo {author} {\bibfnamefont {M.}~\bibnamefont
  {Drescher}}, \bibinfo {author} {\bibfnamefont {M.}~\bibnamefont {Salmhofer}},
  \ and\ \bibinfo {author} {\bibfnamefont {T.}~\bibnamefont {Enss}},\
  }\bibfield  {title} {\bibinfo {title} {\emph {Medium-induced Interaction
  Between Impurities in a Bose-Einstein Condensate}},\ }\href@noop {}
  {\bibfield  {journal} {\bibinfo  {journal} {arXiv preprint arXiv:2303.01916}\
  } (\bibinfo {year} {2023})}\BibitemShut {NoStop}%
\bibitem [{\citenamefont {Camacho-Guardian}\ \emph {et~al.}(2018)\citenamefont
  {Camacho-Guardian}, \citenamefont {Pe\~na Ardila}, \citenamefont {Pohl},\
  and\ \citenamefont {Bruun}}]{Camacho_Guardian2018a}%
  \BibitemOpen
  \bibfield  {author} {\bibinfo {author} {\bibfnamefont {A.}~\bibnamefont
  {Camacho-Guardian}}, \bibinfo {author} {\bibfnamefont {L.~A.}\ \bibnamefont
  {Pe\~na Ardila}}, \bibinfo {author} {\bibfnamefont {T.}~\bibnamefont {Pohl}},
  \ and\ \bibinfo {author} {\bibfnamefont {G.~M.}\ \bibnamefont {Bruun}},\
  }\bibfield  {title} {\bibinfo {title} {\emph {Bipolarons in a Bose-Einstein
  Condensate}},\ }\href {\doibase 10.1103/PhysRevLett.121.013401} {\bibfield
  {journal} {\bibinfo  {journal} {Phys. Rev. Lett.}\ }\textbf {\bibinfo
  {volume} {121}},\ \bibinfo {pages} {013401} (\bibinfo {year}
  {2018})}\BibitemShut {NoStop}%
\bibitem [{\citenamefont {Camacho-Guardian}\ and\ \citenamefont
  {Bruun}(2018)}]{Camacho_Guardian2018b}%
  \BibitemOpen
  \bibfield  {author} {\bibinfo {author} {\bibfnamefont {A.}~\bibnamefont
  {Camacho-Guardian}}\ and\ \bibinfo {author} {\bibfnamefont {G.~M.}\
  \bibnamefont {Bruun}},\ }\bibfield  {title} {\bibinfo {title} {\emph {Landau
  Effective Interaction between Quasiparticles in a Bose-Einstein
  Condensate}},\ }\href {\doibase 10.1103/PhysRevX.8.031042} {\bibfield
  {journal} {\bibinfo  {journal} {Phys. Rev. X}\ }\textbf {\bibinfo {volume}
  {8}},\ \bibinfo {pages} {031042} (\bibinfo {year} {2018})}\BibitemShut
  {NoStop}%
\bibitem [{\citenamefont {Jager}\ and\ \citenamefont
  {Barnett}(2022)}]{Jager_2022}%
  \BibitemOpen
  \bibfield  {author} {\bibinfo {author} {\bibfnamefont {J.}~\bibnamefont
  {Jager}}\ and\ \bibinfo {author} {\bibfnamefont {R.}~\bibnamefont
  {Barnett}},\ }\bibfield  {title} {\bibinfo {title} {\emph {The effect of
  boson–boson interaction on the bipolaron formation}},\ }\href {\doibase
  10.1088/1367-2630/ac9804} {\bibfield  {journal} {\bibinfo  {journal} {New
  Journal of Physics}\ }\textbf {\bibinfo {volume} {24}},\ \bibinfo {pages}
  {103032} (\bibinfo {year} {2022})}\BibitemShut {NoStop}%
\bibitem [{\citenamefont {Ding}\ \emph {et~al.}(2022)\citenamefont {Ding},
  \citenamefont {Drewsen}, \citenamefont {Arlt},\ and\ \citenamefont
  {Bruun}}]{Ding2022}%
  \BibitemOpen
  \bibfield  {author} {\bibinfo {author} {\bibfnamefont {S.}~\bibnamefont
  {Ding}}, \bibinfo {author} {\bibfnamefont {M.}~\bibnamefont {Drewsen}},
  \bibinfo {author} {\bibfnamefont {J.~J.}\ \bibnamefont {Arlt}}, \ and\
  \bibinfo {author} {\bibfnamefont {G.~M.}\ \bibnamefont {Bruun}},\ }\bibfield
  {title} {\bibinfo {title} {\emph {Mediated Interaction between Ions in
  Quantum Degenerate Gases}},\ }\href {\doibase 10.1103/PhysRevLett.129.153401}
  {\bibfield  {journal} {\bibinfo  {journal} {Phys. Rev. Lett.}\ }\textbf
  {\bibinfo {volume} {129}},\ \bibinfo {pages} {153401} (\bibinfo {year}
  {2022})}\BibitemShut {NoStop}%
\bibitem [{\citenamefont {Guenther}\ \emph {et~al.}(2021)\citenamefont
  {Guenther}, \citenamefont {Schmidt}, \citenamefont {Bruun}, \citenamefont
  {Gurarie},\ and\ \citenamefont {Massignan}}]{Guenther2020}%
  \BibitemOpen
  \bibfield  {author} {\bibinfo {author} {\bibfnamefont {N.-E.}\ \bibnamefont
  {Guenther}}, \bibinfo {author} {\bibfnamefont {R.}~\bibnamefont {Schmidt}},
  \bibinfo {author} {\bibfnamefont {G.~M.}\ \bibnamefont {Bruun}}, \bibinfo
  {author} {\bibfnamefont {V.}~\bibnamefont {Gurarie}}, \ and\ \bibinfo
  {author} {\bibfnamefont {P.}~\bibnamefont {Massignan}},\ }\bibfield  {title}
  {\bibinfo {title} {\emph {Mobile impurity in a Bose-Einstein condensate and
  the orthogonality catastrophe}},\ }\href {\doibase
  10.1103/PhysRevA.103.013317} {\bibfield  {journal} {\bibinfo  {journal}
  {Phys. Rev. A}\ }\textbf {\bibinfo {volume} {103}},\ \bibinfo {pages}
  {013317} (\bibinfo {year} {2021})}\BibitemShut {NoStop}%
\end{thebibliography}%

\end{document}